\documentclass[12pt]{article}
\catcode`\@=11
\@addtoreset{equation}{section}

\global\arraycolsep=2pt
\usepackage{mathrsfs,amsbsy,amssymb,latexsym,amsfonts,amsmath,cite}
\usepackage{graphicx,color}
\usepackage{physics}
\usepackage{mathtools}
\usepackage{ulem}

\usepackage{mathrsfs,amsbsy,amssymb,latexsym,amsfonts,amsmath,cite,bm}
\usepackage{graphicx,color}
\usepackage{mathtools}
\usepackage{enumerate}

\newcommand{\no}{\nonumber}

\newcommand{\cA}{\mathcal A}

\newcommand{\cG}{\mathcal G}
\newcommand{\cJ}{\mathcal J}
\newcommand{\cL}{\mathcal L}
\newcommand{\cM}{\mathcal M}

\newcommand{\la}{\lambda}

\newcommand{\llangle}{ \langle\!\langle}
\newcommand{\rrangle}{ \rangle\!\rangle}

\allowdisplaybreaks

\begin{document}

\renewcommand{\thefootnote}{\fnsymbol{footnote}}

\begin{flushright}
KUNS-2847
\end{flushright}
\vspace*{0.5cm}

\begin{center}
{\Large \bf The Faddeev-Reshetikhin model \\[5pt]
\quad from a 4D Chern-Simons theory }
\vspace*{2cm} \\
{\large  Osamu Fukushima$^{\sharp}$\footnote{E-mail:~osamu.f@gauge.scphys.kyoto-u.ac.jp},
Jun-ichi Sakamoto$^{\dagger}$\footnote{E-mail:~sakamoto@ntu.edu.tw},
and Kentaroh Yoshida$^{\sharp}$\footnote{E-mail:~kyoshida@gauge.scphys.kyoto-u.ac.jp}} 
\end{center}

\vspace*{0.4cm}

\begin{center}
$^{\sharp}${\it Department of Physics, Kyoto University, Kyoto 606-8502, Japan.}
\end{center}
\begin{center}
$^{\dagger}${\it Department of Physics and Center for Theoretical Sciences, National Taiwan University, Taipei 10617, Taiwan}
\end{center}

\vspace{2cm}

\begin{abstract}
We derive the Faddeev-Reshetikhin (FR) model from a four-dimensional Chern-Simons theory with two order surface defects by following the work by Costello and Yamazaki 
[arXiv:1908.02289]. Then we present a trigonometric deformation of the FR model 
by employing a boundary condition with an $R$-operator of Drinfeld-Jimbo type. 
This is a generalization of the work by Delduc, Lacroix, Magro and Vicedo 
[arXiv:1909.13824] from the disorder surface defect case to the order one. 
\end{abstract}

\setcounter{footnote}{0}
\setcounter{page}{0}
\thispagestyle{empty}

\newpage

\tableofcontents

\renewcommand\thefootnote{\arabic{footnote}}

\section{Introduction}

Searching for a method to describe various integrable models 
in a unified manner is a significant subject in mathematical physics. 
A nice idea for such a way is to start from four-dimensional gauge theories by following the works  
by Costello, Witten and Yamazaki \cite{Costello:2013zra,CWY1,CWY2}. In particular, two-dimensional (2D) integrable field theories can be derived 
from a four-dimensional Chern-Simons (4D CS) theory 
\begin{align}
    S_{CS}[A]&=\frac{i}{4\pi}\int_{\cM\times C} \omega\wedge \Tr\left[ A \wedge \left(dA+\frac{2}{3}A\wedge A\right)\right] 
\end{align}
equipped with a meromorphic one-form $\omega$ 
\begin{align}
    \omega\equiv\varphi(z)\,dz\, =dz\,,
\end{align}
as proposed by Costello and Yamazaki \cite{CY}.  
The base space is $\cM\times C$\,, where $\cM$ is a 2D manifold and $C$ is a Riemann surface.
Introducing 2D defects enables us to consider a dimensionally reduced theory 
on $\cM$\,. These surface defects are classified into the {\it order} defects  
and the {\it disorder} defects.
The order defects are defined by introducing new degrees of freedom such as 
free fermions and free bosons, which are coupled to the 4D bulk gauge theory.
For the disorder defects, we allow $\omega$ to have zeros on $C$, and the 2D theories lie on the poles of $\omega$.

\medskip

In the disorder defect case, $\omega$ has been identified with a twist function of the associated integrable system \cite{Gaudin}.
Then Delduc, Lacroix, Magro and Vicedo has pushed this perspective and elaborated the procedure to derive integrable field theories 
for disorder defects \cite{DLMV}.
It succeeded in systematically deforming the boundary conditions for $\omega$ with (at most) second-order poles.
Following this procedure, a variety of integrable deformations have been studied \cite{DLMV,FSY1,FSY2,Tian:2020ryu,Tian:2020meg,Hoare:2020mpv,Lacroix:2020flf,Schmidtt:2019otc}. 
However, the order-defect case has not been elaborated so much at least so far. 
For other related works on 4D CS theory, see \cite{Costello:2020lpi,Gaiotto:2020fdr,Gaiotto:2020dhf,Bittleston:2020hfv}.

\medskip

Our puporse here is to discuss the order defect case by focusing upon an example. 
According to the Hamiltonian analysis in \cite{Gaudin}, the models in this case should be ultralocal 
(no $\delta'$-term in the Poisson algebra). A famous example of the ultralocal model is the Faddeev-Reshetikhin model 
\cite{Faddeev:1985qu}. We derive the FR model from a 4D CS theory with two order surface defects. Then we present a trigonometric deformation of the FR model 
by employing a boundary condition with an $R$-operator of Drinfeld-Jimbo type \cite{Drinfeld,Jimbo}. 
This is a generalization of the work \cite{DLMV} from the disorder surface defect case to the order one.

\medskip 

This paper is organized as follows. In section 2, we introduce the basics of the FR model. 
In section 3, the FR model is derived from a 4D Chern-Simons theory with two order surface defects. 
In section 4, we present a trigonometric deformation of the FR model by employing an appropriate 
boundary condition with the $R$-operator of Drinfeld-Jimbo type.  
Section 5 is devoted to conclusion and discussion. 

\medskip

\paragraph{NOTE:} 
Just before completing our draft, we have received an interesting work by Caudrelier, Stoppato and Vicedo \cite{CSV}, where 
the Zakharov-Mikhailov theory (which is a class of ultralocal models) has been derived with order defects \cite{CSV} based 
on the procedure presented in \cite{BSV}. The FR model is included as a special example. But our derivation is different 
from theirs and a trigonometric deformation of it has not been discussed there. 

\newpage 

\section{The Faddeev-Reshetikhin model}

In this section, we shall give a brief review about the Faddeev-Reshetikhin (FR) model \cite{Faddeev:1985qu}.

\subsection{The classical action}

The classical action of the FR model is given by
\begin{align}
    S_{\text{FR}}[g_{(\pm)}]=-\int_{\cM}{\rm Tr}\left(\Lambda g_{(+)}^{-1}\partial_{-} g_{(+)}+\Lambda g_{(-)}^{-1}\partial_{+} g_{(-)}-\frac{1}{2\nu}g_{(+)}\Lambda g_{(+)}^{-1}g_{(-)}\Lambda g_{(-)}^{-1}\right)
    \,d\sigma^+\wedge d\sigma^-\,,\label{FR-action}
\end{align}
where $\nu$ is a real parameter and $g_{(\pm)}$ are group elements of $SU(2)$\,.
Here $\cM$ is 2D Minkowski space with the coordinates 
$x^{\alpha} = (x^0,x^1)=(\tau,\sigma)$ and the metric $\eta_{\alpha\beta} = \mbox{diag}(-1,+1)$\,. 
The light-cone coordinates on $\cM$ are defined as
\begin{align}
\sigma^{\pm}\equiv \frac{1}{2}(\tau\pm\sigma)\,.
\end{align}
Here $\Lambda$ is the Cartan generator of $SU(2)$ taken as
\begin{align}
    \Lambda=T^3\,,
\end{align}
where $T^a\,(a=1,2,3)$ are the generators of $SU(2)$\,,
\begin{align}
    T^a=-\frac{i}{2}\sigma^a\,,\qquad  [T^{a},T^{b}]=\varepsilon^{abc}T^c\,,\qquad \Tr[T^aT^b]=-\frac{1}{2}\delta^{ab}\,.
\end{align}
Here $\sigma^a$ are the Pauli matrices, and the structure constants $\varepsilon^{abc}$ are the antisymmetric tensor normalized 
as $\varepsilon^{123}=1$\,.
The expression (\ref{FR-action}) of the action is given in \cite{Appadu:2017fff}.
The FR model is closely related to the string sigma model with target space $R\times S^3$\,, 
and the low-energy effective action of (\ref{FR-action}) becomes the Landau-Lifshitz model as explained in \cite{Klose:2006dd}.  
It is easy to generalize the action (\ref{FR-action}) to the $SU(N)$ case as discussed in \cite{Appadu:2017fff}, 
but we will restrict ourselves to the $SU(2)$ case for simplicity.

\medskip

The equations of motion obtained from (\ref{FR-action}) are
\begin{align}
    \partial_{\mp}\cJ_{(\pm)}=\mp \frac{1}{2\nu}[\cJ_{(+)},\cJ_{(-)}]\,,\label{FR-eom}
\end{align}
where we have introduced 
\begin{align}
    \cJ_{(\pm)}\equiv g_{(\pm)}\cdot \Lambda \cdot g_{(\pm)}^{-1}\,.\label{cJ-def}
\end{align}
The above equations of motion (\ref{FR-eom}) can be rewritten as
\begin{align}
\partial_+ \cJ_{(-)}-\partial_-\cJ_{(+)}-\frac{1}{\nu}[\cJ_{(+)},\cJ_{(-)}]=0    \,,\qquad \partial_{-}\cJ_{(+)}+\partial_{+}\cJ_{(-)}=0\,.\label{eom-con}
\end{align}
Therefore, $\cJ_{(\alpha)}\,(\alpha=\pm)$ can be regarded as an on-shell conserved current.
While these equations (\ref{eom-con}) have the same forms with the ones derived from the $SU(2)$ PCM, $\cJ_{(\pm)}$ satisfy additional relations 
\begin{align}
    \Tr[(\cJ_{(\pm)})^n]:\text{const}.\label{Jpm-const}
\end{align}
On the other hand, the conserved current of $SU(2)$ PCM does not satisfy this relation.

\medskip

As is well known, the FR model (\ref{FR-action}) is classically integrable. 
Indeed, since the equations (\ref{eom-con}) take the same forms with those for the $SU(2)$ PCM, we can easily construct a Lax pair
\begin{align}
    \cL_{\pm}(z)=\mp\frac{1}{z\pm \nu}\cJ_{(\pm)}\,,\label{FR-lax}
\end{align}
where $z\in\mathbb{C}P^1$ is a spectral parameter.
The flatness condition of the Lax pair (\ref{FR-lax}) is equivalent to the equations of motion (\ref{eom-con}):
\begin{align}
    \partial_+\cL_--\partial_-\cL_++[\cL_+,\cL_-]&=\frac{1}{z-\nu}\partial_+\cJ_{(-)}+\frac{1}{z+\nu}\partial_-\cJ_{(+)}-\frac{1}{z^2-\nu^2}[\cJ_{(+)},\cJ_{(-)}]\no\\
    &=\frac{\nu}{z^2-\nu^2}\left(\partial_+ \cJ_{(-)}-\partial_-\cJ_{(+)}-\frac{1}{\nu}[\cJ_{(+)},\cJ_{(-)}]\right)\no\\
    &\qquad +\frac{z}{z^2-\nu^2}\left(\partial_+ \cJ_{(-)}+\partial_-\cJ_{(+)}\right)\,.
\end{align}
As usual, we can obtain infinite (non-local) conserved charges from the monodromy matrix
\begin{align}
    T(z)={\rm P}\exp\left[-\int_{-\infty}^{\infty}\!\! d\sigma\,\cL_{\sigma}(\sigma;z)\right]\,,
\end{align}
where the symbol ${\rm P}$ denotes the equal-time path ordering in terms of $\sigma$ and the spacial component of the Lax pair is defined as
\begin{align}
    \cL_{\sigma}(\sigma;z)\equiv \frac{1}{2}(\cL_+(\sigma;z)-\cL_{-}(\sigma;z))\,.
\end{align}

\subsection{The Poisson structure}

The Poisson structure of the FR model is much simpler than that of the $SU(2)$ PCM.
In fact, the Poisson brackets of $\cJ_{(\pm)}^{a}(\sigma)$ are given by 
\begin{align}
\begin{split}
    \{\cJ_{(\pm)}^{a}(\sigma_1),\cJ_{(\pm)}^{b}(\sigma_2) \}&=\varepsilon^{abc}\,\cJ_{(\pm)}^{c}(\sigma_2)\delta(\sigma_1-\sigma_2)\,,\\ \{\cJ_{(+)}^{a}(\sigma_1),\cJ_{(-)}^{b}(\sigma_2) \}&=0\,.\label{J-PB}
    \end{split}
\end{align}
These are ultra-local because the term with the derivative of the delta function does not appear in the right hand sides of (\ref{J-PB}), in comparison to the $SU(2)$ PCM.
By using the relations in (\ref{J-PB}), the Poisson bracket of the spacial component of the Lax pair can be expressed as 
\begin{align}
    \{\cL_{\sigma}(\sigma_1;z_1),\cL_{\sigma}(\sigma_2;z_2) \}_{\rm P}&=\left[r(z_1,z_2), \cL_{\sigma}(\sigma_1;z_1)\otimes 1+1\otimes \cL_{\sigma}(\sigma_2;z_2)\right ]\delta(\sigma_1-\sigma_2)\,,\label{PB-Lax}
\end{align}
where the Poisson bracket in the tensorial notation is defined as 
\begin{align}
     \{A,B \}_{\rm P}\equiv \{A\otimes 1,1\otimes B \}=\sum_{a=1}^3\{A^a,B^b\}\,T^a\otimes T^b\,
\end{align}
and $r(z_1,z_{2})\in \mathfrak{g}\otimes \mathfrak{g}$ is the classical $r$-matrix 
associated with the integrable structure of the system. 
The resulting classical $r$-matrix is given by 
\begin{align}
    r(z_1,z_2)=-\frac{\sum_{a=1}^{3} T^a\otimes T^a}{z_1-z_2}=-\frac{\varphi(z_1)^{-1}+\varphi(z_2)^{-1}}{2(z_1-z_2)}\sum_{a=1}^{3} T^a\otimes T^a\,, 
\label{r-rational}
\end{align}
and the twist function $\varphi(z)$ is just one like 
\begin{align}
   \varphi(z)=1\label{FR-twist} \,.
\end{align}
The classical $r$-matrix (\ref{r-rational}) satisfies the classical Yang-Baxter equation (CYBE)
\begin{align}
[r_{12}(z_1,z_2),r_{23}(z_{2},z_3)]+[r_{12}(z_1,z_2),r_{13}(z_{1},z_3)]+[r_{13}(z_1,z_3),r_{23}(z_{2},z_3)]=0\,.\label{CYBE}
\end{align}
Here we have introduced the tensorial notation 
\begin{eqnarray}
r_{12}=r\otimes 1\,,  \quad r_{23}=1\otimes r\,, \quad 
r_{13}=r_{ab}\,(T^a\otimes 1\otimes T^b)\,, 
\end{eqnarray}
where $r_{ab}$ are the components of the $r$-matrix 
\[
r=r_{ab}\,T^a\otimes T^b\,. 
\] 
The relation (\ref{PB-Lax}) leads to the Poisson bracket of the monodromy matrices 
\begin{align}
    \{T(z_1),T(z_2) \}_{\rm P}&=[r(z_1-z_2), (T(z_1)\otimes 1)(1\otimes T(z_2))]\,.
\end{align}
This is the fundamental relation of the FR model.

\section{The FR model from a 4D CS theory}

In this section, we shall derive the FR model from a 4D CS theory 
with two order surface defects. The derivation here is mostly based 
on a generalization of the method in \cite{DLMV} for the disorder case.

\subsection{A 4D CS theory with two order surface defects}

Let us consider a complexified $SU(2)$\,, $G^{\mathbb{C}}=SU(2)^{\mathbb{C}}$\,\footnote{
For consistency with the previous section, we restrict our discussion here to the $G=SU(2)$ case. But the discussion in this section is valid for any semisimple Lie algebra}. The associated complexified Lie algebra is  
$\mathfrak{g}^{\mathbb{C}}\equiv\mathfrak{su}(2)^{\mathbb{C}}$\,.
Then, we consider a $\mathfrak{g}^{\mathbb{C}}$-valued gauge field $A$ defined on 
$\cM \times \mathbb{C}P^1$\,. 
The global holomorphic coordinate of $\mathbb{C}P^1 \equiv\mathbb{C}\cup\{\infty\}$ 
is denoted by $z$\,. This $\mathbb{C}P^1$ geometry characterizes the rational class 
of integrable system. 

\medskip

We start from a 4D CS theory coupled with two order surface defects,
\begin{align}
    S[A,\cG_{(\pm)}]&=S_{CS}[A]-\int_{\cM\times\{z_+\}}\Tr(\Lambda \cdot \cG_{(+)}^{-1}D_{-} \cG_{(+)})\,d\sigma^+\wedge d\sigma^-\no\\
    &\quad-\int_{\cM\times\{z_-\}}
    \Tr(\Lambda\,\cdot\cG_{(-)}^{-1}D_{+} \cG_{(-)})\,d\sigma^+\wedge d\sigma^-\,,\label{action-defect}
\end{align}
where 
the covariant derivatives $D_{\pm}$ are defined as
\begin{align}
    D_{\pm} \cG_{(\mp)}\equiv (\partial_{\pm}+\cA_{\pm})\cG_{(\mp)}\,,\qquad 
    \cA_{+}\equiv A_{+}|_{z_{-}}\,,\quad \cA_{-}\equiv A_{-}|_{z_{+}}\,.
\end{align}
The second and third terms of (\ref{action-defect}) describe the two order surface defects sitting at $z_{\pm}\in \mathbb{R}$\,, respectively.
The first term is the 4D CS action given by 
\begin{align}
    S_{CS}[A]&=\frac{i}{4\pi}\int_{\cM\times \mathbb{C}P^1} \omega\wedge CS(A)\,,\label{4dcs}
\end{align}
where $CS(A)$ is the CS three-form defined as 
\begin{align}
CS(A)\equiv \Tr\left[ A \wedge \left(dA+\frac{2}{3}A\wedge A\right)\right]\,.
\end{align}
Here, the meromorphic one-form $\omega$ is defined in terms of the twist function (\ref{FR-twist}) as
\begin{align}
    \omega\equiv\varphi(z)\,dz\, =dz\,,
\end{align}
which has a double pole
\begin{align}
    \mathfrak{p}=\{\infty\}\,.
\end{align}
Note here that since $\omega$ is a (1,0)-form, the action (\ref{4dcs}) has an extra gauge symmetry
\begin{align}
A\mapsto A+\chi\, dz\,.\label{extra gauge}
\end{align}
It enables us to take the gauge $A_z=0$\,, i.\,e., 
\begin{align}
A=A_\sigma\, d\sigma+A_\tau\, d\tau+A_{\bar{z}}\, d\bar{z}\,.
\end{align}

\subsubsection*{Equations of motion}

Let us derive the equations of motion of the action (\ref{action-defect}).
Taking a variation of (\ref{action-defect}) with respect to $A$, we obtain
\begin{align}
    \delta S[A]&=\frac{i}{2\pi}\int_{\cM\times \mathbb{C}P^1} \omega\wedge \Tr\left(  \delta A \wedge F(A)\right) +
    \frac{i}{4\pi}\int_{\cM\times \mathbb{C}P^1} d\omega\wedge \Tr\left(  \delta A\wedge A\right)\no\\
  &\quad- \int_{\cM\times \mathbb{C}P^1} {\rm Tr}(\delta A_{-}\cdot\cG_{(+)} \Lambda\, \cG_{(+)}^{-1}\,\delta(z-z_+))\,d\sigma^+\wedge d\sigma^-\wedge dz \wedge d \bar{z}\no\\
    &\quad -\int_{\cM\times \mathbb{C}P^1}
   {\rm Tr}(\delta A_{+}\cdot \cG_{(-)} \Lambda\,\cG_{(-)}^{-1}\,\delta(z-z_-))\,d\sigma^+\wedge d\sigma^-\wedge dz \wedge d \bar{z} \,,
   \label{eq:action-var}
\end{align}
where $F(A)\equiv d A+A\wedge A$ is the field strength of $A$\,.
Here, we have assumed that $A$ vanishes at the boundary of $\cM\times \mathbb{C}P^1$, and used the relation of the delta function
\begin{align}
   \int_{\mathbb{C}P^1} \delta(z-z_{\pm})\, dz\wedge d\bar{z}=1\,.
\end{align}
Then, the bulk equations of motion are given by
\begin{align}
   F_{+-}&=0\,,\label{eq:beom1}\\
 \omega\,F_{\bar{z}+}&=-2\pi i\,\cG_{(+)}\cdot \Lambda\cdot\cG_{(+)}^{-1}\,\delta(z-z_+)dz\,,\label{eq:beom2}\\
  \omega\,F_{\bar{z}-}&=2\pi i\,\cG_{(-)}\cdot\Lambda\cdot \cG_{(-)}^{-1}\,\delta(z-z_-)dz \,. \label{eq:beom3}
\end{align}
The second and third equations indicate that $A$ has poles at $z=z_{\pm}$\,.
For later discussion, we denote the set of the positions of the order surface defects as 
\begin{align}
    \mathfrak{z}=\{z_{\pm}\}\,.
\end{align}
It is useful to rewrite the boundary equation of motion as 
\begin{align}
   & (\text{res}_{\infty}\omega)\epsilon^{\alpha\beta}\Tr\left( A_{\alpha}\lvert_{\infty}\,\delta A_{\beta}\lvert_{\infty} \right)
    +(\text{res}_{\infty}\xi_\infty\omega)\epsilon^{\alpha\beta}\partial_{\xi_\infty}\Tr\left( A_{\alpha}\,\delta A_{\beta} \right)\lvert_{\infty}\no\\
    =& -2\epsilon^{\alpha\beta}\partial_{\xi_\infty}\Tr\left( A_{\alpha}\,\delta A_{\beta} \right)\lvert_{\infty} =0\,,\label{bd-eom}
\end{align}
where $\xi_{\infty}\equiv 1/z$ is the local coordinate around $z=\infty$\,.

\subsubsection*{Gauge invariance}

Let us see here the gauge invariance of the action (\ref{4dcs})\,.  

\medskip

In analogy with the disorder defect case \cite{DLMV}, it is natural to consider 
a gauge transformation 
\begin{align}
A\mapsto A^u\equiv u\cdot A\cdot u^{-1}-duu^{-1}\,,  \qquad
\cG_{(\pm)}\mapsto \cG_{(\pm)}^{u}\equiv u|_{z_{\pm}}\cdot \cG_{(\pm)}\,,
\label{gauge transformation}
\end{align}
where $u$ is a $G^{\mathbb{C}}$-valued function defined on 
$\cM\times\mathbb{C}P^1$\,. 
Then, at the off-shell level, the action (\ref{action-defect}) is transformed under the transformations 
(\ref{gauge transformation}) 
as
\begin{align}
S[A^u]=S[A]+\frac{i}{4\pi}\int_{\cM\times \mathbb{C}P^1} \omega \wedge I_{WZ}[u]+ \frac{i}{4\pi}\int_{\cM\times \mathbb{C}P^1} \omega\wedge d\left(\Tr\left( u^{-1}du\wedge A\right)\right)\,, 
\label{gauge variation}
\end{align}
where $I_{WZ}[u]$ is the Wess-Zumino (WZ) three-form defined as
\begin{align}
I_{WZ}[u]\equiv\frac{1}{3}\Tr( u^{-1}du\wedge u^{-1}du\wedge u^{-1}du)\,.
\end{align}
Thus the action (\ref{action-defect}) is invariant if the gauge parameter $u$ satisfies
\begin{align}
\frac{i}{4\pi}\int_{\cM\times \mathbb{C}P^1} \omega \wedge I_{WZ}[u]=0\,,
\qquad u|_{\mathfrak{p}}=1\,. \label{u-cond}
\end{align}
These conditions are the same as in the disorder defect case. As a result, the transformations (\ref{gauge transformation}) 
can be regarded as a gauge transformation with $u$ satisfying the condition (\ref{u-cond})\,.



\subsection{Lax form}\label{subsection:Lax-form}

Let us next introduce a Lax pair associated with the action (\ref{action-defect}).

\medskip

As in the disorder defect case, a Lax pair is introduced by performing a formal gauge transformation\footnote{For the terminology ``formal''\,, 
see \cite{DLMV,FSY1}.}  (\ref{gauge transformation}),
\begin{align}
   A&=-d\hat{g}\hat{g}^{-1}+\hat{g}\cdot \cL\cdot \hat{g}^{-1}\,,\qquad 
 \cG_{(\pm)}=\hat{g}_{(\pm)}\cdot g_{(\pm)}\,,
    \label{A-L}
\end{align}
where $\hat{g}\,, g_{(\pm)}\in G^{\mathbb{C}}$ and $\hat{g}_{(\pm)}\equiv \hat{g}|_{z_{\pm}}$\,.
Here, we take a gauge choice such that $\cL_{\bar{z}}=0$\,, and hence the one-form $\cL$ takes the form
\begin{align}
    \cL=\cL_{\tau}d\tau+\cL_{\sigma}d\sigma=\cL_{+}d\sigma^++\cL_{-}d\sigma^-\,.\label{L-form0}
\end{align}
By substituting (\ref{L-form0}) into (\ref{eq:beom2}), (\ref{eq:beom3}), the bulk equations of motion become
\begin{align}
  &\partial_+\cL_--\partial_-\cL_++[\cL_+,\cL_-]=0\,,\label{eq:Lbeom1}\\
  &\omega\,\partial_{\bar{z}}\cL_{+}=-2\pi i\,\cJ_{(+)}\,\delta(z-z_+)dz\,,\label{eq:Lbeom2}\\
  &\omega\,\partial_{\bar{z}}\cL_{-}=2\pi i\,\cJ_{(-)}\,\delta(z-z_-)dz \,. \label{eq:Lbeom3}
\end{align}
The currents $\cJ_{(\pm)}$ are defined as 
\begin{align}
\cJ_{(\pm)}\equiv g_{(\pm)}\cdot\Lambda\cdot g_{(\pm)}^{-1}\,.
\end{align}
As we will see later, these are going to be identified with the current (\ref{cJ-def}).
The boundary equations of motion (\ref{eq:Lbeom2}) and (\ref{eq:Lbeom3}) indicate that the Lax pair is a $\mathfrak{g}^{\mathbb{C}}$-valued meromorphic one-form with poles $z=z_{\pm}$\,.

\medskip

By substituting (\ref{A-L}) into (\ref{action-defect}), the 4D action (\ref{action-defect}) can be written as
\begin{align}
    S[A]&=\frac{i}{4\pi}\int_{\cM\times \mathbb{C}P^1}\omega \wedge\left(\Tr(\cL\wedge d\cL)+ d\left(\Tr(\hat{g}^{-1}d\hat{g}\wedge \cL)\right)+ I_{WZ}[\hat{g}]\right)\no\\
    &\quad -\int_{\cM\times\{z_+\}}\Tr\left(\Lambda\cdot g_{(+)}^{-1}(\partial_{-}+\cL_-|_{z_+}) g_{(+)}\right)\,d\sigma^+\wedge d\sigma^-\no\\
    &\quad -\int_{\cM\times\{z_-\}}
    \Tr\left(\Lambda\cdot g_{(-)}^{-1}(\partial_{+}+\cL_+|_{z_-}) g_{(-)}\right)\,d\sigma^+\wedge d\sigma^-\,.\label{4D-FR-action}
\end{align}
Note that the expression (\ref{4D-FR-action}) is still a 4D action.
In the next subsection, we will dimensionally reduce the 4D action (\ref{4D-FR-action}) 
to the corresponding 2D action by imposing conditions on $\hat{g}$\,.

\subsection{From 4D to 2D via the archipelago conditions}

In order to obtain the associated 2D integrable model, it is necessary to impose the archipelago conditions \cite{DLMV} on $\hat{g}$ as in the disorder defect case.
Then, the 4D action (\ref{4dcs}) is drastically simplified as follows:
\begin{align}
    S[g_{(\pm)}]&=-\frac{1}{4}\sum_{x\in \mathfrak{p}}\int_{\cM }\Tr\left(\text{res}_{x}(\varphi\,\cL\right)\wedge \hat{g}^{-1}_{x}d\hat{g}_{x})-\frac{1}{4}\sum_{x\in \mathfrak{p}}\int_{\cM\times [0,R_x]}(\text{res}_{x}\,\omega) \wedge I_{WZ}[\hat{g}_x]\no\\
    &\quad +\frac{i}{4\pi}\int_{\cM\times \mathbb{C}P^1}\omega \wedge \Tr(\cL\wedge d\cL)\no\\
    &\quad -\int_{\cM}\Tr\left(\Lambda\,g_{(+)}^{-1}\partial_{-}g_{(+)}+\Lambda\, g_{(-)}^{-1}\partial_{+}g_{(-)}+\cJ_{(+)} \cL_-|_{z_+}+\cJ_{(-)} \cL_{+}|_{z_-}\right)\,d\sigma^+\wedge d\sigma^-\,,\no\\
    &=
    +\frac{i}{4\pi}\int_{\cM\times \mathbb{C}P^1}\omega \wedge \Tr(\cL\wedge d\cL)\no\\
    &\quad -\int_{\cM}\Tr\left(\Lambda\,g_{(+)}^{-1}\partial_{-}g_{(+)}+\Lambda\, g_{(-)}^{-1}\partial_{+}g_{(-)}+\cJ_{(+)} \cL_-|_{z_+}+\cJ_{(-)} \cL_+|_{z_-}\right)\,d\sigma^+\wedge d\sigma^-\,,
    \label{effaction}
\end{align}
where in the second equality we have used the relations
\begin{align}
&\text{res}_\infty\,(\varphi\,\cL)=0\,,\qquad\text{res}_\infty\,\omega=0\,.
\end{align}
The integrand of the first term in (\ref{effaction}) is apparently a four-form, but as we will see in (\ref{LdL-formula}) it is localized on the defects at $\mathcal{M}\times \{z_{\pm}\}$ because $d\cL$ in the integrand generates delta functions due to the bulk equations of motion (\ref{eq:Lbeom2}), (\ref{eq:Lbeom3}).

\subsubsection*{Reality condition}
Let us now discuss the reality condition to ensure that the 4D action (\ref{action-defect}) is real.
An involution $\mu_{\rm t}:\mathbb{C}P^{1}\to\mathbb{C}P^1$ is defined by complex conjugation $z\mapsto\bar{z}$\,.
Let $\tau:\mathfrak{g}^{\mathbb{C}}\to\mathfrak{g}^{\mathbb{C}}$ be an anti-linear involution, and then the set of the fixed points under $\tau$ defines a real subalgebra $\mathfrak{g}$ of $\mathfrak{g}^{\mathbb{C}}$\,.
The involutive automorphism $\tau$ satisfies 
\begin{align}
\overline{\Tr(B\wedge C)}=\Tr(\tau B\wedge \tau C)\,,\qquad {}^{\forall}B,C\in\mathfrak{g}^{\mathbb{C}}\,.
\end{align}
The associated operation on the Lie group $G^{\mathbb{C}}$ is denoted by $\tilde{\tau}:G^{\mathbb{C}}\to G^{\mathbb{C}}$\,.

\medskip

The reality condition is imposed through these involutions as
\begin{align}
\bar{\omega}=\mu_{\rm t}^{*}\omega\,,\qquad 
\tau A=\mu_{\rm t}^{*}A\,,\qquad
\tilde{\tau}\cG_{(\pm)}=\mu_{\rm t}^{*}\cG_{(\pm)}\,.
\label{reality-condition}
\end{align}
One can see that the action (\ref{action-defect}) is real under the condition (\ref{reality-condition}):
\begin{align}
\overline{S[A,\cG_{(\pm)}]}
=&\,
-\frac{i}{4\pi}\int_{\cM\times\mathbb{C}P^{1}}\bar{\omega}\wedge CS(\tau A)\no\\
& -\int_{\cM\times\{z_+\}}\Tr\left(\Lambda\cdot \tilde{\tau}\cG_{(+)}^{-1}(\partial_{-}+\tau A_-|_{z_+}) \tilde{\tau}\cG_{(+)}\right)\,d\sigma^+\wedge d\sigma^-\no\\
    & -\int_{\cM\times\{z_-\}}
    \Tr\left(\Lambda\cdot \tilde{\tau}\cG_{(-)}^{-1}(\partial_{+}+\tau A_+|_{z_-}) \tilde{\tau}\cG_{(-)}\right)\,d\sigma^+\wedge d\sigma^-\no\\
=&\,
-\frac{i}{4\pi}\int_{\cM\times \mathbb{C}P^{1}}\mu_{\rm t}^{*}\omega\wedge CS(\mu_{\rm t}^{*}A) \no\\
& -\int_{\cM\times\{z_+\}}\Tr\left(\Lambda\cdot \mu_{\rm t}^{*}\cG_{(+)}^{-1}(\partial_{-}+ \mu_{\rm t}^{*}A_-|_{z_+}) \mu_{\rm t}^{*}\cG_{(+)}\right)\,d\sigma^+\wedge d\sigma^-\no\\
    & -\int_{\cM\times\{z_-\}}
    \Tr\left(\Lambda\cdot \mu_{\rm t}^{*}\cG_{(-)}^{-1}(\partial_{+}+ \mu_{\rm t}^{*}A_+|_{z_-}) \mu_{\rm t}^{*}\cG_{(-)}\right)\,d\sigma^+\wedge d\sigma^-\no\\
=&\,
-\frac{i}{4\pi}\int_{\cM\times \mu_{\rm t}\mathbb{C}P^{1}}\omega\wedge CS(A) \no\\
& -\int_{\cM\times\mu_{\rm t}\{z_+\}}\Tr\left(\Lambda\cdot \cG_{(+)}^{-1}(\partial_{-}+ A_-|_{z_+}) \cG_{(+)}\right)\,d\sigma^+\wedge d\sigma^-\no\\
    & -\int_{\cM\times\mu_{\rm t}\{z_-\}}
    \Tr\left(\Lambda\cdot \cG_{(-)}^{-1}(\partial_{+}+ A_+|_{z_-}) \cG_{(-)}\right)\,d\sigma^+\wedge d\sigma^-\no\\
=&\,S[A,\cG_{(\pm)}]\,.
\end{align}
Note here that $\mu_{\rm t}(z_{\pm})=z_{\pm}$ and $\Lambda\in\mathfrak{g}$\,. 
From the relation (\ref{A-L}), the reality condition is also expressed as
\begin{align}
\tilde{\tau}\hat{g}_{(\pm)}=\mu_{\rm t}^{*}\hat{g}_{(\pm)}\,,\quad
\tilde{\tau}g_{(\pm)}=\mu_{\rm t}^{*}g_{(\pm)}\,,\quad
\tau\cL=\mu_{\rm t}^{*}\cL\,.
\label{reality-Lax}
\end{align}

\subsubsection*{2D gauge symmetry}
The 2D action (\ref{effaction}) has the ``2D gauge symmetry''.
One can perform the 2D gauge transformations keeping $A$ and $\cG_{(\pm)}$ unchanged and preserving the archipelago conditions imposed on $\hat{g}$\,.
Under the transformation, $\cL$\,, $\hat{g}_{(\pm)}$ and $g_{(\pm)}$ are transformed as
\begin{align}
\cL\mapsto h^{-1}dh+h^{-1}\cdot \cL \cdot h\,,\qquad
\hat{g}_{(\pm)}\mapsto \hat{g}_{(\pm)}\cdot h\,,\qquad
g_{(\pm)}\mapsto h^{-1}\cdot g_{(\pm)}\,,
\label{2D gauge}
\end{align}
where $h$ is a smooth $\mathfrak{g}$-valued function depending on $(\tau,\sigma)\in\cM$\,.
In contrast to the 4D gauge symmetry (\ref{gauge transformation}), 
the 2D gauge symmetry (\ref{2D gauge}) is considered as the redundancy in defining $\hat{g}$ without altering $A$ and $\cG_{(\pm)}$\,.

\subsubsection*{2D effective action}

In order to evaluate (\ref{effaction}), let us determine the explicit expression 
of the Lax form.

\medskip

The first is to solve the boundary equation of motion (\ref{bd-eom}) with the following condition:
\begin{align}
    A\lvert_{\infty}=0\,.\label{boundary-dirichlet}
\end{align}
This is a trivial solution to the boundary equation of motion.

\medskip

As we saw in section \ref{subsection:Lax-form}, $\cL_{\pm}$ have poles at $z=z_{\pm}$\,, respectively.
Therefore, it is natural to suppose the following form of $\cL_{\pm}$:
\begin{align}
    \cL=\left(U_{+}-\frac{\cJ_{(+)}}{z-z_+}\right)d\sigma^++\left(U_{-}+\frac{\cJ_{(-)}}{z-z_-}\right)d\sigma^-\,,\label{L-ansatz}
\end{align}
where we have used a formula for delta functions
\begin{align}
    \delta(z-z_{\pm})=\frac{1}{2\pi i}\frac{\partial}{\partial \bar{z}}\left(\frac{1}{z-z_{\pm}}\right)\,.
\end{align}
Here $U_{\pm}$ are undetermined functions on $\cM$ and take values in $\mathfrak{g}$ due to the reality condition (\ref{reality-Lax}).
The 2D gauge symmetry (\ref{2D gauge}) allows us to set an archipelago type field $\hat{g}$ like
\begin{align}
 \hat{g}\lvert_{\infty}=1\,.
\end{align}
Since the boundary condition (\ref{boundary-dirichlet}) indicates that $U_{\pm}=0$\,, 
the Lax form is determined as
\begin{align}
    \cL=-\frac{\cJ_{(+)}}{z-z_+}d\sigma^++\frac{\cJ_{(-)}}{z-z_-}d\sigma^-\,.
    \label{LaxFR-4D}
\end{align}

\medskip

Finally, let us evaluate the 4D action (\ref{effaction}).
By using the expression (\ref{LaxFR-4D}), the first term in (\ref{effaction}) 
can be rewritten as
\begin{align}
     &-\frac{i}{4\pi}\int_{\cM\times \mathbb{C}P^1}\omega \wedge \Tr(\cL\wedge d\cL)\no\\
     &=\frac{1}{2}\int_{\cM\times \mathbb{C}P^1} \Tr\left[\cL_+\left(\cJ_{(-)}\delta(z-z_-)\right)\right]\,dz\wedge d\sigma^+\wedge d\bar{z}\wedge d\sigma^-\no\\
     &\qquad+\Tr\left[\cL_-\left(-\cJ_{(+)}\delta(z-z_+)\right)\right]\,d z \wedge d\sigma^-\wedge d \bar{z}\wedge d \sigma^+\no\\
     &=\frac{1}{2}\int_{\cM} \Tr\left(\cJ_{(-)}\cL_+|_{z_-}+\cJ_{(+)}\cL_-|_{z_+}\right)\,d\sigma^+\wedge d\sigma^-\,.\label{LdL-formula}
\end{align}
Then, the 2D effective action is evaluated as 
\begin{align} 
& S_{\rm 2D}[g_{(\pm)}] \notag \\ 
    =&-\int_{\cM}\Tr\left(\Lambda\,g_{(+)}^{-1}\partial_{-}g_{(+)}+\Lambda\, g_{(-)}^{-1}\partial_{+}g_{(-)}+\frac{1}{2}\cJ_{(+)} \cL_-\big|_{z_+}+\frac{1}{2}\cJ_{(-)} \cL_+\big|_{z_-}\right)\,d\sigma^+\wedge d\sigma^-\no\\
    =&-\int_{\cM}\Tr\left(\Lambda\,g_{(+)}^{-1}\partial_{-}g_{(+)}+\Lambda\, g_{(-)}^{-1}\partial_{+}g_{(-)}+\frac{1}{(z_+-z_-)}\cJ_{(+)} \cJ_{(-)}\right)\,d\sigma^+\wedge d\sigma^-\,.\label{actionFR-4D}
\end{align}
The above expressions (\ref{LaxFR-4D}), (\ref{actionFR-4D}) agree with (\ref{FR-lax}) and (\ref{FR-action}) if we take $z_+=-\nu$ and $z_-=\nu$\,.  
In this sense, the resulting action (\ref{actionFR-4D}) is a slightly generalized version of the original FR model.

\section{A trigonometric-deformation of the FR model}

In this section, let us consider a trigonometric-deformation of the FR model.

\subsection{A twist function}

For this purpose, we replace the rational classical $r$-matrix (\ref{r-rational}) with the $\mathfrak{su}(2)$ trigonometric $r$-matrix\footnote{For the $\mathfrak{sl}(2)$ case, see \cite{CWY1,CY}.}\,,
\begin{align}
    r_{\text{trig.}}(\la_1,\la_2)=\frac{ i\eta\,T^+\otimes T^-}{1-e^{i\eta\,(\la_1-\la_2)}}-\frac{ i\eta\,T^-\otimes T^+}{1-e^{-i\eta\,(\la_1-\la_2)}}-\frac{\eta}{2}\cot\left(\frac{\eta(\la_1-\la_2)}{2}\right)T^3\otimes T^3\,,\label{tri-r-la}
\end{align}
where we have introduced a deformation parameter $\eta\in \mathbb{R}$ and 
\begin{align}
    T^{\pm}=\frac{1}{\sqrt{2}}(T^1\pm i T^2)\,.
\end{align}
Note that the classical $r$-matrix (\ref{tri-r-la}) satisfies the CYBE (\ref{CYBE}).
The spectral parameter $\la$ takes a value on a cylinder (rather than $\mathbb{C}P^1$) 
because the classical $r$-matrix (\ref{tri-r-la}) is of trigonometric type. Then the fundamental region of $\la$ is represented by 
\begin{align}
    \mathbb{C}/\mathbb{Z}=\left\{~\la\in\mathbb{C}~\biggl\lvert~ -\frac{\pi}{2\eta}< \Re \la <\frac{3\pi}{2\eta}~\right\}\,.
\end{align}
By taking a limit $\eta \to 0$\,, the classical $r$-matrix (\ref{tri-r-la}) reduces to the rational one (\ref{r-rational}).
Note that the $r$-matrix (\ref{tri-r-la}) is skew-symmetric in terms of spectral parameters and its components,
\begin{align}
    r_{\text{trig.}ab}(\la_1,\la_2)=-r_{\text{trig.}ba}(\la_2,\la_1)\,.
\end{align}
Here we have defined the components of the $r$-matrix as
\begin{align}
    r_{\text{trig.}}(\la_1,\la_2)\equiv r_{\text{trig.},ab}(\la_1,\la_2)\,T^a\otimes T^b\,.
\end{align}
Since the associated twist function is obtained as a measure of the asymmetry of a given classical $r$-matrix, the $(1,0)$-form $\omega$ should be taken as
\begin{align}
    \omega=\varphi_{\text{trig.}}(\la)\,d \la=d\la\,.\label{la-omega}
\end{align}

\subsubsection*{A relationship with Costello and Yamazaki }
It is instructive to see that 
the choice (\ref{la-omega}) of $\omega$ is consistent with the expression in \cite{CY}.

\medskip 

First, let us move from the cylinder $\mathbb{C}/\mathbb{Z}$ to a plane $\mathbb{C}^\times=\mathbb{C}\backslash\{0\}$ via the map
\begin{align}
   z=e^{i\eta\,\la}\,.\label{tr-z-la}
\end{align}
Note that in the $z$-coordinate system, 
the trigonometric $r$-matrix (\ref{tri-r-la}) becomes 
\begin{align}
    r_{\text{trig.}}(z_1,z_2)=\frac{ i\eta}{1-z_1/z_2}T^+\otimes T^--\frac{ i\eta}{1-z_2/z_1}T^-\otimes T^+-\frac{ i\eta}{2}\frac{ z_1+z_2}{z_1-z_2}T^3\otimes T^3\,.  
\label{tri-r}
\end{align}
This is related to the rational one (\ref{r-rational}) through the relation
\begin{align}
   r_{\text{trig.}}(z_1,z_2)\equiv \frac{\varphi^{-1}_{\text{trig.}}(z_1)+\varphi^{-1}_{\text{trig.}}(z_2)}{2}r(z_1,z_2) \,.\label{twist-def}
\end{align}

\medskip 

Then, the $(1,0)$-form $\omega$ on $\mathbb{C}^{\times}$ takes the form 
\begin{align}
    \omega= \varphi_{\text{trig.}}(z)\,dz=\frac{dz}{i\eta\,z}\,,\label{tri-omega}
\end{align}
and has two simple poles
\begin{align}
    \mathfrak{p}=\{0,\infty\}\,.\label{pole-tri}
\end{align}
The form of $\omega$ in (\ref{tri-omega}) is the same as in the one in \cite{CY}.

\subsubsection*{The reality condition of $\omega$}

An involution $\mu_{\rm t}$ may be defined as follows:
\begin{align}
    \mu_{\rm t}:~ \la \to\overline{\la} \quad\Longleftrightarrow \quad z\to \frac{1}{\overline{z}}\,. 
    \label{involution-trig}
\end{align}
In the $\la$ coordinate, the reality condition is trivial:
\begin{align}
    \overline{\omega}=d\overline{\la}=\mu_{\rm t}\omega\,.\label{real-omega-la}
\end{align}

\subsection{A boundary condition}

In the following, we will consider the 4D CS action (\ref{action-defect}) 
with the (1,0)-form (\ref{tri-omega}).
We obtain the same bulk equations of motion (\ref{eq:beom1}), (\ref{eq:beom2}) and (\ref{eq:beom3}), but now $\omega$ is replaced by the one in (\ref{tri-omega}).
Note that in this section, the order surface defects lie on $z=z_{\pm}$ such that $z_{\pm}=1/\bar{z}_{\pm}$ since the involution $\mu_{\rm t}$ is defined as (\ref{involution-trig}).

\medskip

The $(1,0)$-form $\omega$ has the two simple poles (\ref{pole-tri}).
Hence, the boundary equations of motion are 
\begin{align}
   &(\text{res}_{0}\,\omega)\epsilon^{\alpha\beta}\Tr\left( A_{\alpha}\lvert_{0}\,\delta A_{\beta}\lvert_{0} \right)+ (\text{res}_{\infty}\,\omega)\epsilon^{\alpha\beta}\Tr\left( A_{\alpha}\lvert_{\infty}\,\delta A_{\beta}\lvert_{\infty} \right)
   \no\\
    =& \epsilon^{\alpha\beta} \llangle\left(A_\alpha|_0,A_\alpha|_{\infty}\right),\delta \left(A_\beta|_0,A_\beta|_{\infty}\right) \rrangle =0\,,\label{q-bd-eom}
\end{align}
where the bilinear form is defined as
\begin{align}
 \llangle\left(x,x'\right),\left(y,y'\right) \rrangle\equiv\frac{1}{i\eta}\left(\Tr\left( x\cdot x'\right)-\Tr\left(y\cdot y'\right)\right)\,.
\end{align}
As shown in \cite{FSY1}, the boundary condition (\ref{q-bd-eom}) can be solved by assigning the following Drinfeld double to the bilinear form 
\begin{align}
\mathfrak{h}\equiv
\mathfrak{g}^\delta\oplus\mathfrak{g}_{R}\,,
\label{def:d=gd+gR}
\end{align}
where $\mathfrak{g}^\delta$ and $\mathfrak{g}_R$ are defined as 
\begin{align}
\mathfrak{g}_R&\equiv \left\{((R-i)x,(R+i)x)|x\in\mathfrak{g}\right\}\,,\label{def:gR}\\
\mathfrak{g}^\delta&\equiv \left\{(x,x)|x\in\mathfrak{g}\right\}\,.\label{def:gd}
\end{align}
Here, $R:\mathfrak{g}\to \mathfrak{g}$ is a skew-symmetric $R$-operator satisfying the modified classical Yang-Baxter equation (mCYBE)
\begin{align}
[R(x),R(y)]-R\left([R(x),y]+[x,R(y)]\right)=[x,y] \qquad
(x,y\in\mathfrak{g},\; R\in \operatorname{End}\mathfrak{g})\,,
\label{mCYBE}
\end{align}
and 
\begin{align}
\Tr\left( R(x)y \right)=-\Tr\left(x R(y)\right)\,, \qquad \forall x,y\in\mathfrak{g}\,.
\end{align}
Here, let us take the $R$-operator of the Drinfeld-Jimbo type  \cite{Drinfeld,Jimbo} 
such that
\begin{align}
R(T^{\pm})=\mp i T^{\pm},\qquad R(T^3)=0\,.
\label{eq:R-su(2)}
\end{align}
We can easily check that the $R$-operator satisfies the mCYBE (\ref{mCYBE}).

\medskip

As a result, $A_{\alpha}$ is supposed to satisfy
\begin{alignat}{2}
(A_{\alpha}|_{0},A_{\alpha}|_{\infty})\in \mathfrak{g}_{R}\,. 
\label{manin0}
\end{alignat}

\subsection{The associated Lax form and 2D action}

Let us next derive the associated Lax form and 2D action. 

\medskip

As in the rational case, we can easily see that the associated Lax form satisfies the equations (\ref{eq:Lbeom1}), (\ref{eq:Lbeom2}) and (\ref{eq:Lbeom3}) though 
$\omega$ is now replaced with (\ref{tri-omega}).
Hence, an ansatz of $\cL_{\pm}$ is taken as
\begin{align}
    \cL=\left(U_{+}-\frac{i\eta\,z\,\cJ_{(+)}}{z-z_+}\right)d\sigma^++\left(U_{-}+\frac{i\eta\,z\,\cJ_{(-)}}{z-z_-}\right)d\sigma^-\,,\label{qL-ansatz}
\end{align}
where $U_{\pm}$ are undetermined smooth functions $\cM \to \mathfrak{g}^{\mathbb{C}}$\,.
The reality condition is again realized as in (\ref{reality-Lax})\,.

\medskip

In order to obtain the expression of $U_{\pm}$\,, we will take boundary conditions 
as in (\ref{manin0}).
Then, the constraints on $A_{\pm}$ are given by
\begin{align}
    (R-i)A_{\pm}\lvert_{0}=(R+i)A_{\pm}\lvert_{\infty}\,.\label{A-qFR-beom}
\end{align}
Since the choice of the Drinfeld double (\ref{def:d=gd+gR}) enable us to take $\hat{g}\lvert_{z=0}\in G$, one can take $\hat{g}\lvert_{z=0}=1$ by using the 2D gauge invariance under $g \to g \cdot h\, (h\in G)$\,. 
Furthermore, the condition $\tilde{\tau}\hat{g}=\mu_t^*\hat{g}$ indicates $\tilde{\tau}(\hat{g}\lvert_{z=0})=\mu^*_t(\hat{g}\lvert_{z=0})=\hat{g}\lvert_{z=\infty}$\,. 
Then by using the gauge symmetry, we can take
\begin{align}
    \hat{g}\lvert_{z=0}=\hat{g}\lvert_{z=\infty}=1\,.
\end{align}
Then, the constraints (\ref{A-qFR-beom}) become
\begin{align}
(R-i)U_+=(R+i)(U_+- i\eta\cJ_{(+)})\,,\qquad (R-i)U_-=(R+i)(U_-+ i\eta\cJ_{(-)})\,.
\end{align}
By solving the equations, we obtain
\begin{align}
    U_{\pm}=\pm\frac{\eta}{2}(R+i)\cJ_{(\pm)}\,.
\end{align}
Therefore, the resulting Lax form is given by 
\begin{align}
    \cL=\left(\frac{\eta}{2}(R+i)-\frac{i\eta\,z}{z-z_+}\right)\cJ_{(+)}d\sigma^++\left(-\frac{\eta}{2}(R+i)+\frac{i\eta\,z}{z-z_-}\right)\cJ_{(-)}d\sigma^-\,,
    \label{LaxrFR-4D}
\end{align}
which indeed satisfies the reality condition (\ref{reality-Lax}):
\begin{align}
\tau \cL=&\,
\left(\frac{\eta}{2}(R-i)-\frac{-i\eta\,\bar{z}}{\bar{z}-\bar{z}_+}\right)\tau\cJ_{(+)}d\sigma^++\left(-\frac{\eta}{2}(R-i)+\frac{-i\eta\,\bar{z}}{\bar{z}-\bar{z}_-}\right)\tau\cJ_{(-)}d\sigma^-\no\\
=&\,
\left(\frac{\eta}{2}(R-i)-\frac{-i\eta\,\bar{z}}{\bar{z}-z_+^{-1}}\right)\cJ_{(+)}d\sigma^++\left(-\frac{\eta}{2}(R-i)+\frac{-i\eta\,\bar{z}}{\bar{z}-z_-^{-1}}\right)\cJ_{(-)}d\sigma^-\no\\
=&\,
\mu_{\rm t}^{*}\left[\left(\frac{\eta}{2}(R-i)-\frac{-i\eta\,z^{-1}}{z^{-1}-z_+^{-1}}\right)\cJ_{(+)}d\sigma^++\left(-\frac{\eta}{2}(R-i)+\frac{-i\eta\,z^{-1}}{z^{-1}-z_-^{-1}}\right)\cJ_{(-)}d\sigma^-\right]\no\\
=&\,
\mu_{\rm t}^{*}\left[\left(\frac{\eta}{2}(R+i)-\frac{i\eta\,z}{z-z_+}\right)\cJ_{(+)}d\sigma^++\left(-\frac{\eta}{2}(R+i)+\frac{i\eta\,z}{z-z_-}\right)\cJ_{(-)}d\sigma^-\right]\no\\
=&\,\mu_{\rm t}^{*}\cL\,.
\end{align}
Here we have used the fact that $\cJ_{(\pm)}$ take values in the real Lie algebra $\mathfrak{g}$\,, and $z_{\pm}\in \mathbb{C}^\times$ satisfy the condition $z_{\pm}=1/\bar{z}_{\pm}$\,.
More interestingly, the Lax form (\ref{LaxrFR-4D}) can be expressed in terms of the trigonometric $r$-matrix (\ref{tri-r}).
To see this, let us expand the current $\cJ_{(\pm)}$ as 
\begin{align}
\cJ_{(\pm)}=\cJ_{(\pm)}^-T^++\cJ_{(\pm)}^+T^-+\cJ_{(\pm)}^3T^3\,,
\end{align}
and then the Lax pair (\ref{LaxrFR-4D}) can be rewritten as
\begin{align}
    \cL&=
\left(-\frac{i\eta\,z_+}{z-z_+}\cJ_{(+)}^-T^+-\frac{i\eta\, z}{z-z_+}\cJ_{(+)}^+T^-+\frac{i\eta}{2}\frac{z+z_+}{z-z_+}\cJ_{(+)}^3T^3\right)d\sigma^+\no\\
&\quad+\left(\frac{i\eta\,z_-}{z-z_-}\cJ_{(-)}^-T^++\frac{i\eta\, z}{z-z_-}\cJ_{(-)}^+T^--\frac{i\eta}{2}\frac{z+z_-}{z-z_-}\cJ_{(-)}^3T^3\right)d\sigma^-\no\\
&=\left(\sum_{a=\pm,3}r_{\text{trig.},ab}(z,z_+)\cJ_{(+)}^bT^a\right)d\sigma^+
+\left(-\sum_{a=\pm,3}r_{\text{trig.},ab}(z,z_-)\cJ_{(-)}^bT^a\right)d\sigma^-\,.\label{Lax-r}
\end{align}
This expression (\ref{Lax-r}) takes a similar form presented in \cite{CY}.

\medskip

Finally, let us derive the associated 2D action.
As in the rational case, we can use the same formula (\ref{LdL-formula}) though $\mathbb{C}P^1$ in (\ref{LdL-formula}) is replaced with $\mathbb{C}^{\times}$.
As a result, the resulting 2D action is given by
\begin{align}
    &S_{2D}[g_{(\pm)}]\no\\
    =&-\int_{\cM}\Tr\left(\Lambda\,g_{(+)}^{-1}\partial_{-}g_{(+)}+\Lambda\, g_{(-)}^{-1}\partial_{+}g_{(-)}+\frac{1}{2}\cJ_{(+)} \cL_-\big|_{z_+}+\frac{1}{2}\cJ_{(-)} \cL_+\big|_{z_-}\right)\,d\sigma^+\wedge d\sigma^-\no\\
    =& -\int_{\cM}\Tr\biggl(\Lambda\,g_{(+)}^{-1}\partial_{-}g_{(+)}+\Lambda\, g_{(-)}^{-1}\partial_{+}g_{(-)}+\frac{i\eta}{2}\frac{z_++z_-}{z_+-z_-}\cJ_{(+)} \cJ_{(-)}-\frac{\eta}{2}\cJ_{(+)}R(\cJ_{(-)})\biggr)\,d\sigma^+\wedge d\sigma^-\no\\
=&-\int_{\cM}\Tr\biggl(\Lambda\,g_{(+)}^{-1}\partial_{-}g_{(+)}+\Lambda\, g_{(-)}^{-1}\partial_{+}g_{(-)}+\frac{\eta\,\cJ_{(+)} \cJ_{(-)}}{2\tan\left(\frac{\eta(\la_+-\la_-)}{2}\right)}-\frac{\eta}{2}\cJ_{(+)}R(\cJ_{(-)})\biggr)\,d\sigma^+\wedge d\sigma^-\,,\label{actionqFR-4D}
\end{align}
where we have parametrized the positions $z_{\pm}\in \mathbb{C}^\times$ of the defects as 
\begin{align}
z_{\pm}=\exp(i\eta\,\la_{\pm})\,,\qquad \la_{\pm}\in\mathbb{R}\,.
\end{align}
The deformed action (\ref{actionqFR-4D}) can also be expressed in terms of the trigonometric $r$-matrix,
\begin{align}
    &S_{2D}[g_{(\pm)}]\no\\
    =& -\int_{\cM}\biggl(\Tr\Bigl(\Lambda\,g_{(+)}^{-1}\partial_{-}g_{(+)}+\Lambda\, g_{(-)}^{-1}\partial_{+}g_{(-)}\Bigr)+r_{\text{trig.},ab}(\la_+,\la_-)\cJ_{(-)}^{a}\cJ_{(+)}^{b}\biggr)\,d\sigma^+\wedge d\sigma^-\,.\label{actionqFR-4D-r}
\end{align}
Note that by taking a limit $\eta \to 0$\,, the 2D action (\ref{actionqFR-4D}) reduces to the undeformed one (\ref{actionFR-4D}).

\section{Conclusion and Discussion}

In this paper, we have derived the FR model from a 4D CS theory with two order surface defects. 
Then we have presented a trigonometric deformation of the FR model by employing the boundary condition 
with the $R$-operator of Drinfeld-Jimbo type. 
This is a generalization of the work \cite{DLMV} from the disorder surface defect case to the order one. 

\medskip 

There are open questions. It is well known that a lattice regularized model exists 
for the FR model \cite{Faddeev:1985qu}.
The integrable lattice model should be realized by considering the expectation value of 
Wilson lines in 4D CS theory \cite{CWY1,CWY2}.
It would be important to understand how the continuum limit of the expectation value gives the 4D CS action (\ref{action-defect}) associated with the FR model, as described in Figure 1 of \cite{CY}.
In relation to this issue, it would also be interesting to see how the quantum inverse scattering method 
can be applied at the level of 4D CS theory. 

\medskip

Moreover, as discussed in \cite{CostelloYagi}, integrable lattice models can be realized by considering brane configurations. Hence, the lattice model associated with the FR model should also be described 
by a certain brane configuration. In particular, it would be interesting to understand the brane description of the FR model by taking its continuous limit.

\subsection*{Acknowledgments}

We would like to thank H.~Y.~Chen and B.~Vicedo for useful discussions 
during the online workshop on ``Online 2020 NTU-Kyoto high energy physics workshop.''
The work of J.S.\ was supported in part by Ministry of Science and Technology (project no. 109-2811-M-002-539), 
National Taiwan University.
The works of K.Y.\ was supported by the Supporting Program for Interaction-based Initiative Team Studies (SPIRITS) from Kyoto University, and JSPS Grant-in-Aid for Scientific Research (B) No.\,18H01214. This work is also supported in part by the JSPS Japan-Russia Research Cooperative Program.



\end{document}